\documentclass[10pt,twocolumn,letterpaper]{article}

\usepackage{iccv}
\usepackage{times}
\usepackage{epsfig}
\usepackage{graphicx}
\usepackage{amsmath}
\usepackage{algorithm,algorithmic}
\usepackage{amssymb}
\usepackage{subfigure}

\usepackage[numbers,sort&compress,square]{natbib}

\DeclareMathOperator*{\argmin}{arg\,min}

\usepackage[pagebackref=true,breaklinks=true,colorlinks,bookmarks=false]{hyperref}

\iccvfinalcopy 

\def\iccvPaperID{31} 

\ificcvfinal\fi

\begin{document}

\title{Integrating Data and Image Domain Deep Learning for Limited Angle Tomography using Consensus Equilibrium}

\author{Muhammad Usman Ghani, W. Clem Karl\\
Boston University\\
{\tt\small \{mughani, wckarl\}@bu.edu}
}

\maketitle
\ificcvfinal\thispagestyle{empty}\fi

\begin{abstract}
Computed Tomography (CT) is a non-invasive imaging modality with applications ranging from healthcare to security. It reconstructs cross-sectional images of an object using a collection of projection data collected at different angles. Conventional methods, such as FBP, require that the projection data be uniformly acquired over the complete angular range. In some applications, it is not possible to acquire such data. Security is one such domain where non-rotational scanning configurations are being developed which violate the complete data assumption. Conventional methods produce images from such data that are filled with artifacts. The recent success of deep learning (DL) methods has inspired researchers to post-process these artifact laden images using deep neural networks (DNNs). This approach has seen limited success on real CT problems. Another approach has been to pre-process the incomplete data using DNNs aiming to avoid the creation of artifacts altogether. Due to imperfections in the learning process, this approach can still leave perceptible residual artifacts. In this work, we aim to combine the power of deep learning in both the data and image domains through a two-step process based on the consensus equilibrium (CE) framework. Specifically, we use conditional generative adversarial networks (cGANs) in both the data and the image domain for enhanced performance and efficient computation and combine them through a consensus process. We demonstrate the effectiveness of our approach on a real security CT dataset for a challenging $90^0$ limited-angle problem. The same framework can be applied to other limited data problems arising in applications such as electron microscopy, non-destructive evaluation, and medical imaging.
\end{abstract}

\section{Introduction}

\begin{figure}
    \centering
    \includegraphics[width=0.5\textwidth]{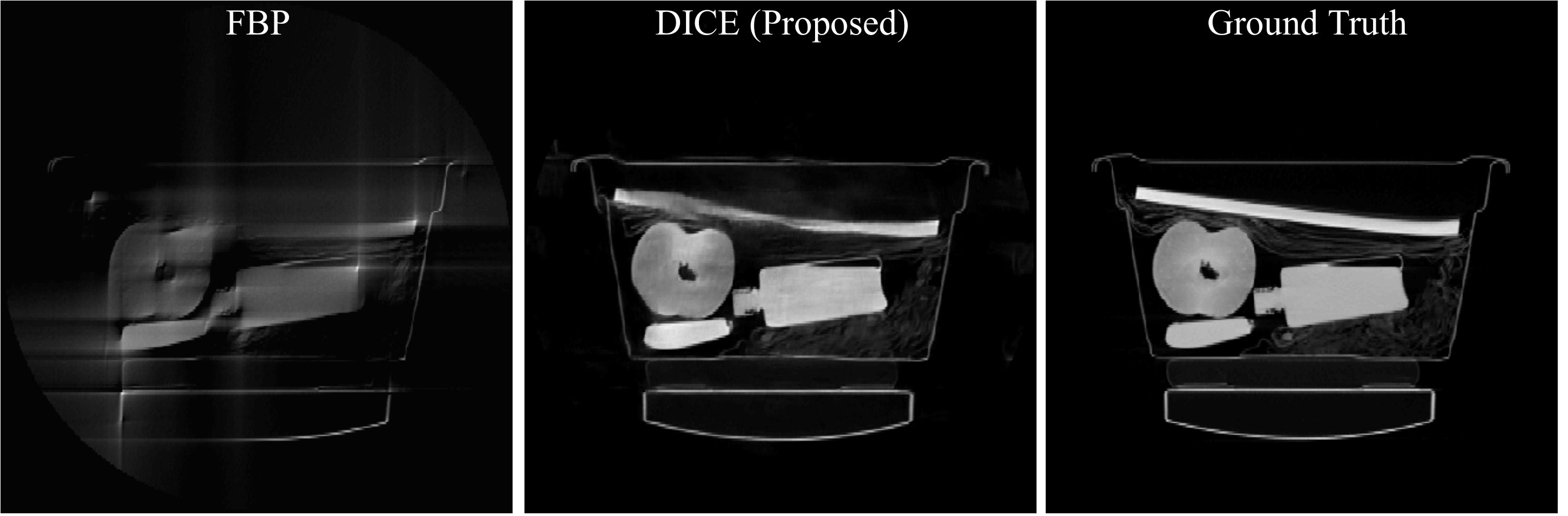}
    \caption{FBP and proposed DICE method reconstructions using $90^0$ limited angle data are presented along with a reference full-view MBIR reconstruction.}
    \label{fig:fbp}
\end{figure}

$X$-ray computed tomography (CT) is a fundamental imaging tool for many applications in medical healthcare \cite{elbakri2002statistical,chang2017modeling}, materials science \cite{tuysuzoglu2016variable,mohan2015timbir}, industrial testing \cite{carmignato2018industrial}, and security \cite{martin2015learning,ghani2018correction}. It reconstructs cross-sectional images of objects by collecting a series of projection data at all angles and processing the acquired projection data using an image reconstruction algorithm. In a typical 2-D parallel-beam data acquisition setup, data is acquired at all angles $\theta \in [0^0, 180^0]$. The filtered backprojection algorithm (FBP) is the most widely used algorithm for CT image reconstruction due to its simple and efficient implementation. An alternative approach is model-based image reconstruction (MBIR) \cite{jin2015model}, which allows incorporation of both a physical model and prior information about the objects being imaged. MBIR can reconstruct higher quality images as compared to FBP, but its iterative nature makes it computationally very expensive, which has limited its adoption. In certain situations, it becomes impractical to acquire data with full angular coverage, which creates a limited angle CT problem. Conventional, computationally efficient methods such as FBP produce artifact-filled images from such limited angle CT data (see Figure \ref{fig:fbp}), with the severity of the artifacts generally becoming more prominent the more limited the data becomes. \par

In this paper, we propose to use the consensus equilibrium framework (CE) \cite{buzzard2018plug} to integrate prior information from both the data and the image domains along with knowledge of the $X$-ray physics for efficient and improved limited angle CT reconstructions. We encapsulate the prior information from the data and the image domains using deep neural networks (DNNs). In contrast to existing image post-processing-only approaches, our data domain deep learning (DL) component learns to complete the projection data and therefore aims to avoid the creation of image domain artifacts. Our image domain DNN learns to improve image quality by learning patch-based priors. Unlike typical image priors such as Total-Variation (TV) \cite{ritschl2011improved} or Markov Random Field (MRF) \cite{zhang2016gaussian}, our image domain DNN efficiently learns the salient image features from a large dataset. Moreover, it has been shown in \cite{ye2018deep,peuizzari2019coherent} that image domain DNN priors can lead to faster convergence. Both of our DNNs are trained using a conditional generative adversarial networks (cGAN) strategy \cite{pix2pix17}. Overall our proposed method aim to combine the quality benefits of MBIR methods with the flexibility and efficiency of DL methods, combining the data and image domain DL via consensus equilibrium (CE) framework.



\subsection{Motivation and Challenges}
Limited angle CT scans can arise due to the physical limitations of the data acquisition or to achieve other desirable imaging capabilities. For example, acquiring data for a limited number of views could significantly reduce the scan time. This would allow CT imaging of highly dynamic organs in the body such as the heart \cite{cho2013motion} without introduction of blurring. This would also enable CT based study of physical processes in dynamic objects for materials science applications \cite{mohan2015timbir,tuysuzoglu2017fast}. It could also be useful for tomographic imaging of specific regions of interest such as for dental \cite{arai2001practical} or breast \cite{niklason1997digital} scan. Our own motivation has been checkpoint security where new, non-rotating gantry configurations use a limited number of static $X$-ray sources resulting in a limited angle CT image formation problem \cite{tuysuzoglu2016variable}. \par



Deep learning has had great impact in image enhancement in the computer vision community, where the presence of powerful and flexible network models coupled with large datasets \cite{krizhevsky2012imagenet} and efficient and inexpensive GPU-based computational resources for training have resulted in impressive processing results \cite{xie2012image,kohler2014mask,kim2016accurate,zhang2017beyond,pix2pix17,zhao2017loss,chaudhury2017can}. Motivated by such successes researchers in the tomographic community have rushed to apply these tools to restore the artifact filled images produced by conventional CT reconstruction methods such as FBP applied to reduced quality CT data. While these DL methods have greatly enhanced CT reconstruction, they fail to completely solve the problem, especially in severely limited data cases. Relative to medical problem, the problem is even more challenging in the security application, where the underlying scene can have a much larger range of shapes, objects, and materials. 



\subsection{Contributions}
The main contributions of this work are summarized below.

\begin{itemize}
    \item Presentation of a framework to combine data domain and image domain DL in CT image reconstruction.
    \item Demonstration of the potential of combined data and image DL for limited angle CT arising in security problem that outperforms existing post-processing methods.
    \item Example images and performance metrics on a real security dataset \cite{crawford2014advances}. 
\end{itemize}

\section{Related Work}

Data-driven models have become increasingly popular in image reconstruction research in recent years, including for limited angle CT applications. A recent survey paper by Ravishankar \etal\cite{ravishankar2019image} compactly summarizes these advances. Image post-processing using data-derived learned models has been the most popular theme, where a DNN is trained on images directly \cite{zhang2016image} or in the wavelet domain \cite{gu2017multi} with the aim of enhancing low-quality FBP-derived images. While these methods can effectively enhance FBP images, they can still fail to recover image features completely; a task which becomes even more challenging in security settings. An alternative approach is to treat the problem in the data domain using a DNN which learns to complete the projection data \cite{li2019sinogram}. However, incorporating projection data consistency conditions is very difficult in DL frameworks \cite{anirudh2017lose}. Anirudh \etal proposed an end-to-end DL framework that learns to reconstruct images directly from limited angle data \cite{anirudh2017lose}. This approach fails to incorporate the $X$-ray physical model in the learning framework and uses a fully connected layer with a huge number of parameters in the generator network. Wurfl \etal \cite{wurfl2018deep} developed a DNN architecture inspired from FBP that learns to adjust projection domain weights to enhance images. This method however has limited flexibility and therefore results in only minor improvements.\par

A related problem is sparse-view CT which has been widely studied. Fewer projection data are acquired with uniform spacing over complete angular range. Similar approaches have been explored for this sparse-view CT: image post-processing \cite{jin2017deep,yang2017low,li2017low,han2018framing}, projection data completion \cite{lee2018deep,ghani2018deep,dong2019sinogram}, combination of projection data completion and image post-processing \cite{liang2018comparision}, and end-to-end learning \cite{zhu2018image,he2019radon}.\par

Another popular theme is regularized inversion with implicitly or explicitly defined priors. The Plug-and-Play priors approach \cite{venkatakrishnan2013plug,sreehari2016plug} does not require priors to be explicitly defined, and therefore allows easy integration of image domain DNNs in a regularized inversion framework \cite{ye2018deep}. Similar ideas have been explored using other formulations where variable splitting and replacement of proximal operators by learned alternatives is performed \cite{meinhardt2017learning,ono2017primal,kamilov2017plug,gupta2018cnn}. The RED method \cite{romano2017little} is a similar approach except that it explicitly defines the prior term. \par

\section{Integrated Data and Image Domain Deep Learning}

\begin{figure}
    \centering
    \includegraphics[width=0.4\textwidth]{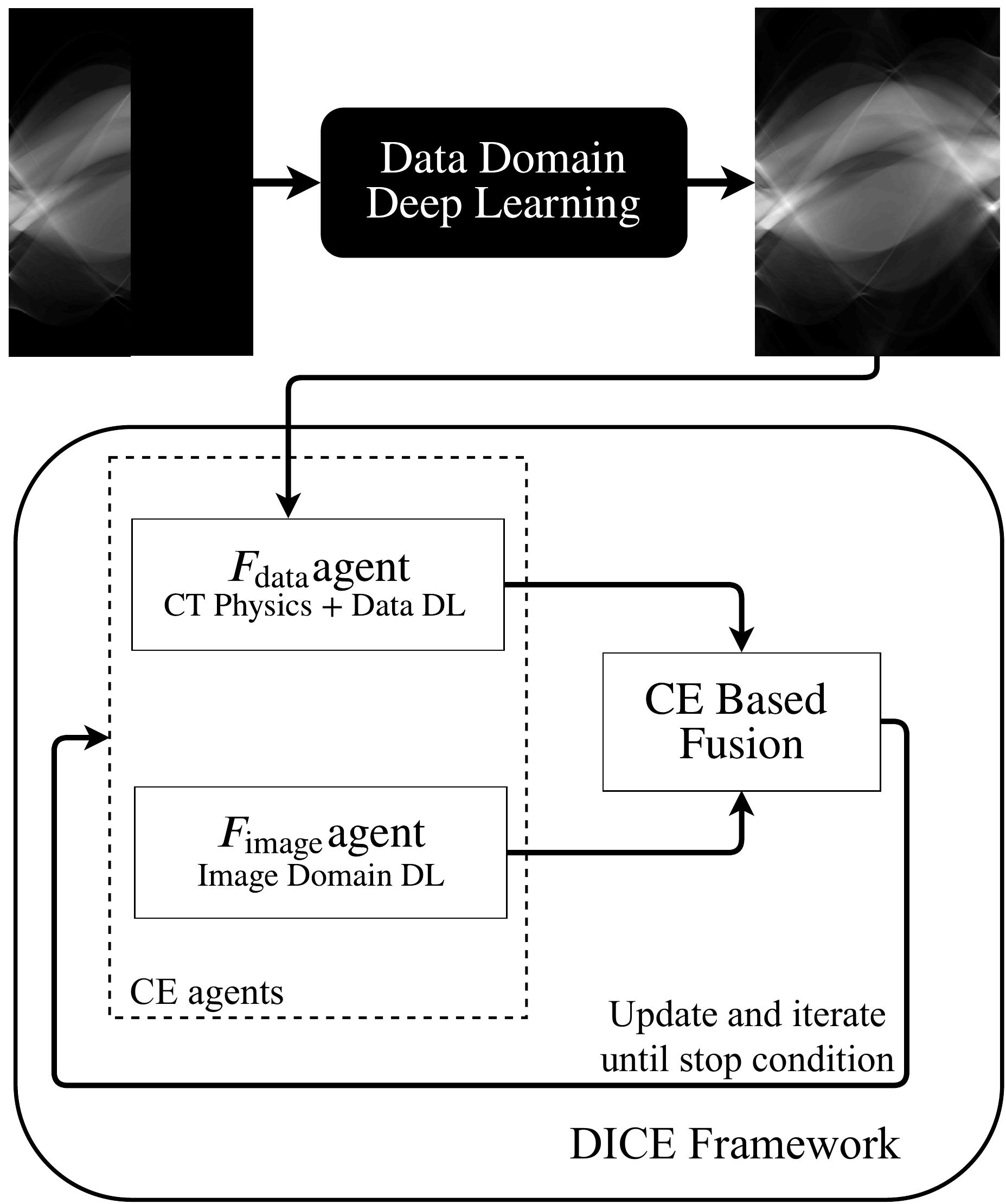}\\
    \caption{Overview of our DICE framework is presented. Data and image domain DL is integrated into CT image reconstruction using consensus equilibrium framework.}
    \label{fig:dice}
\end{figure}

Consensus equilibrium framework \cite{buzzard2018plug,Bouman:18} was developed as a generalization of the ADMM-derived plug-and-play approach \cite{venkatakrishnan2013plug,sreehari2016plug} as a principled means to integrate multiple heterogenous models or ``agents" to yield a single coherent reconstruction. CE starts with a collection of models or agents and derives a consensus solution to the collection. In our case there will be two such agents and the CE equations would be:
\begin{eqnarray}
	F_{\textnormal{data}} (v_1^* = x^* + u_1^*) & = & x^*  \\
	F_{\textnormal{image}}(v_2^* = x^* + u_2^*)  & = & x^*  \\
	\mu_1 u_1^*+ \mu_2 u_2^*  & = & 0  
\end{eqnarray}
where $x^*=\mu_1 v_1^*+ \mu_2 v_2^*$ is the CE estimate of $x$, where $\mu_i$ defines the individual contribution of each agent, and $\mathbf{u}$ can be interpreted as noise vector. Intuitively, the functions $F$ are chosen to map initial values of $x$ to improved values and the solution to the CE equations determines a set of inputs $x_k$ that balance the forces from the competing agents. See \cite{buzzard2018plug,Bouman:18} for a detailed discussion. 

In this work, we combine data and image domain DL for CT image reconstruction using CE and call it DICE framework. An overview of our DICE framework is presented in Figure \ref{fig:dice} with major components of our approach. We have two CE agents build around two DL models. The first block combines data domain DL model with CT physics. The model attempts to complete the limited projection data which is embedded into CT physical model. The second block uses image domain DL and is focused on enhancing tomographic images. The impact of both blocks is combined through the consensus equilibrium (CE) framework, which provides a rational way of combining information from multiple sources. In particular, we construct a CE formulation consisting of two terms. $F_\textnormal{data}$ is CT-physics-derived data consistency constraint and $F_\textnormal{image}$ is the DL image domain model. The data consistency term is based on completed data from the data domain DL model. In this way both data domain and image domain DL models are combined for an improved overall outcome. In the following sections we describe each of these pieces.

\subsection{Data Domain Deep Learning}

\begin{figure*}
    \centering
    \includegraphics[width=0.9\textwidth]{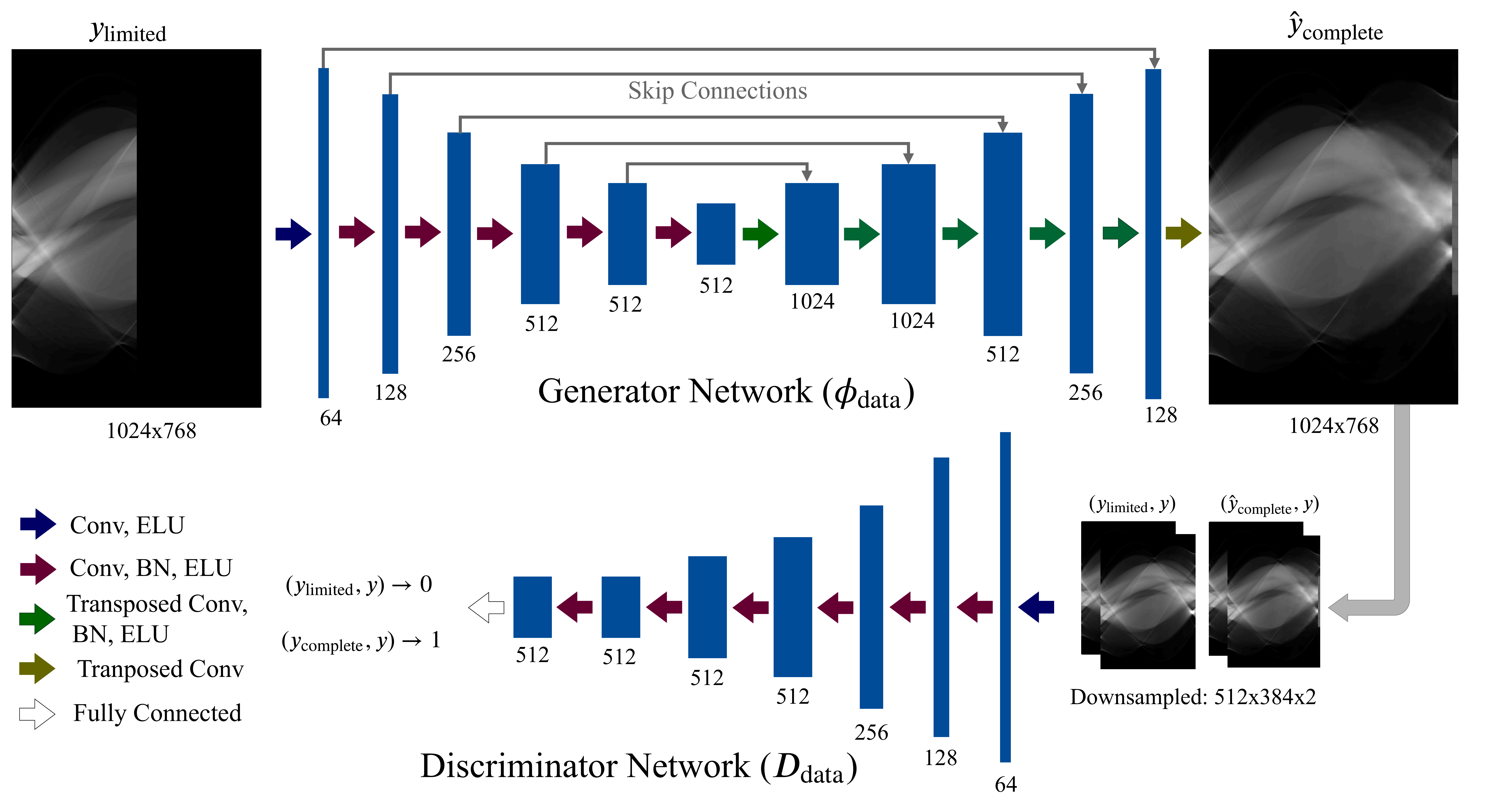}
    \caption{Overall architecture of our data completion cGAN is presented. The abbreviated legends in the Figure are defined here; Conv: 2D convolution, ELU: exponential linear unit, BN: batch-normalization, and 2D Transposed Conv: transposed convolution.  }
    \label{fig:cgan_unet}
\end{figure*}

In this section we describe the data domain DL block in Figure \ref{fig:dice}. This block is inspired from data domain DL work by Ghani and Karl \cite{ghani2018deep}. The data domain DL block performs the following steps in sequence:
\begin{equation}
	\begin{aligned}
		\hat{y}_{\textnormal{complete}} &= \phi_{\textnormal{data}}(y_{\textnormal{limited}})\\
		\hat{x} &= \text{FBP}(\hat{y}_{\textnormal{complete}})\\
		\hat{y}_{\textnormal{consistent}} &= \mathbf{A} \hat{x} \\
	\end{aligned}
	\label{eq:preproc}
\end{equation}
where $y_{\textnormal{limited}} \in \mathbb{R}^{M/2}$ is the observed limited angle data, $\phi_{\textnormal{data}}$ is the data domain DNN trained to complete the limited data, $\hat{y}_{\textnormal{complete}} \in \mathbb{R}^{M}$ is the corresponding estimated complete projection data, $\mathbf{A} \in \mathbb{R}^{M \times N}$ is the tomographic forward projection operator, and $\hat{y}_{\textnormal{consistent}} \in \mathbb{R}^{M}$ is the estimated consistent and completed projection data.\par

Valid tomographic projection data must satisfy a set of consistency conditions \cite{prince1990constrained}. In general, $\hat{y}_{\textnormal{complete}}$, obtained as the output of our trained DNN, will not satisfy these consistency conditions. While we could simply use this inconsistent data estimate in the following processing, we instead choose to run this projection estimate through an inversion and re-projection cycle to obtain the consistent projection data in $\hat{y}_{\textnormal{consistent}}$. Experience has shown this imposition of data consistency results in improved overall results. Direct mapping of $\hat{y}_{\textnormal{complete}}$ onto the space of consistent projection data is the focus of future work. \par

The key operation in (\ref{eq:preproc}) is the DNN mapping represented by $\phi_{\textnormal{data}}$. We use a cGAN for our data-domain DNN, which is composed of a generator network which performs data completion coupled with a discriminator network.  Isola \etal \cite{pix2pix17} reported that using such a combination results in better network performance. The generator and discriminator networks in our cGAN is jointly trained by optimizing a mini-max cost function \cite{pix2pix17,goodfellow2014generative}. The overall architecture of our data completion cGAN is presented in Figure \ref{fig:cgan_unet}. Our generator network ($\phi_{\textnormal{data}}$) and discriminator ($D_{\textnormal{data}}$) both are inspired from \cite{pix2pix17,ghani2019deepmar}. We use $7\times7$ kernels in the convolutional and transposed convolutional layers in both networks. The convolution and transponsed convolution operations are performed with a $2$-pixel stride in both networks. The number of channels used at each layer are given at the bottom of each layer output. The generator network $\phi_{\textnormal{data}}$ has a fully convolutional architecture. It has $6$ down-sampling and $6$ up-sampling layers. The down-sampling layers use $2$-pixel strided convolutions, and the up-sampling layers use $2$-pixel strided transponsed convolutions. In addition, the skip connections are used to transport and concatenate high-resolution information from the down-sampling layers to the up-sampling layers. The generator $\phi_{\textnormal{data}}$ has a theoretical effective receptive field (ERF) of $1135\times1135$. It is trained to perform blind projection data completion for efficient training, i.e., $\phi_{\textnormal{data}}(y_{\textnormal{limited}}) = y_{12}$, where $y_{12}$ is output of last layer in $\phi_{\textnormal{data}}$. At test time, we perform mask-specific data completion, i.e., $\phi_{\textnormal{data}}(y_{\textnormal{limited}}) = y_{\textnormal{limited}} + M \odot y_{12}$, where $M$ is the mask representing missing views and $\odot$ is an element-wise multiplication operation. The discriminator network, $D_{\textnormal{data}}$, has $7$ convolutional and a fully connected layer. For efficient training, we use down-sampled projection data as input to $D_{\textnormal{data}}$.

\subsection{Image Domain Deep Learning}

\begin{figure*}
	\centering
	\includegraphics[width=0.7\textwidth]{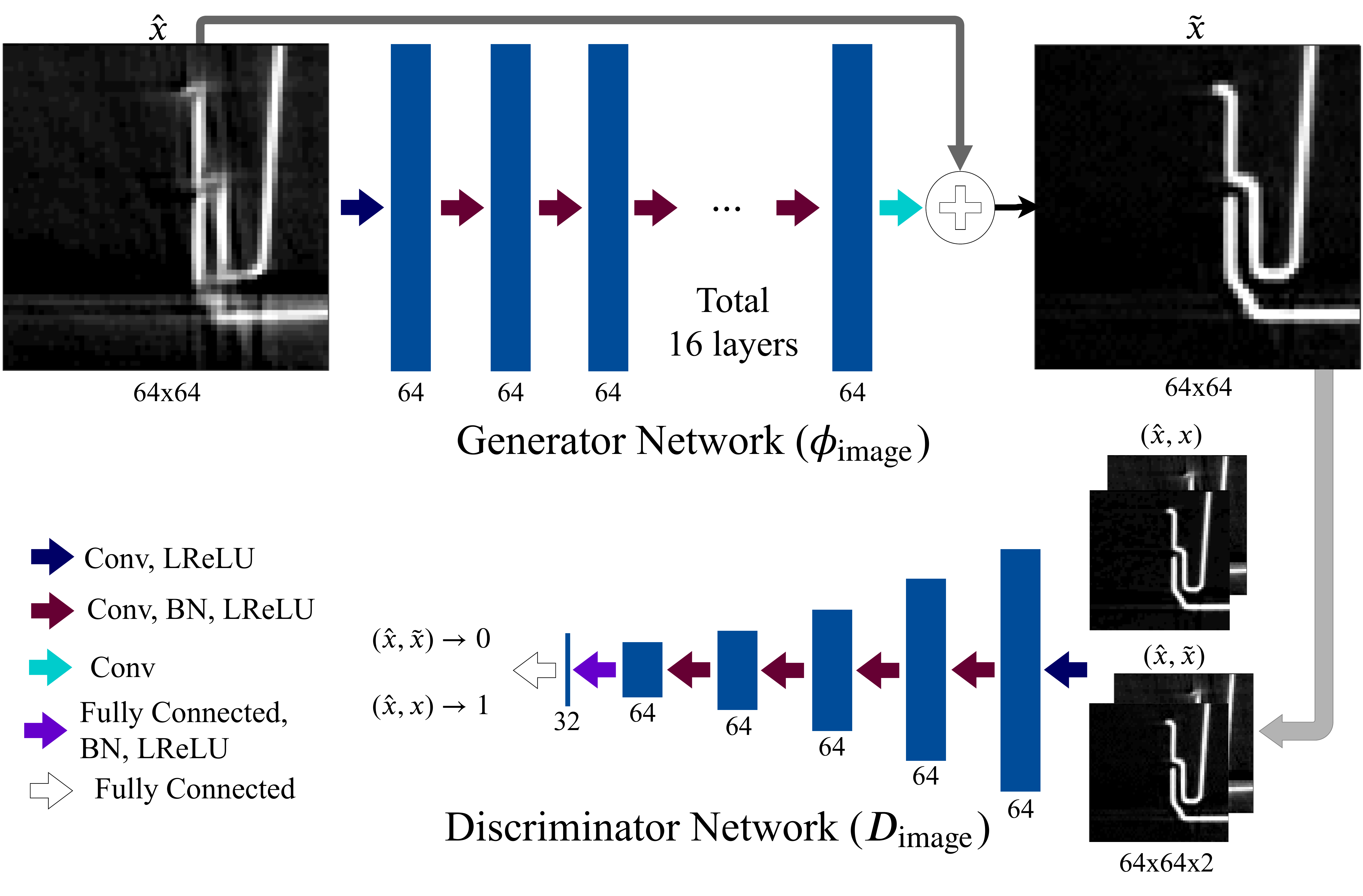}
	\caption{Overall architecture of our image-domain cGAN is presented. It learns patch-based image priors from a large security dataset. The abbreviated terms in the Figure are defined here; Conv: 2D convolution, LReLU: leaky rectified linear unit, and BN: batch-normalization. }
	\label{fig:cgan_res_vdsr}
\end{figure*}

In this section we describe the image domain DL block in Figure~\ref{fig:dice}. Similar to our data-domain DNN, we use a cGAN for $\phi_{\textnormal{image}}$, which is composed of a generator network which performs image enhancement coupled with a discriminator network.  The generator and discriminator networks in our cGAN is jointly trained as described earlier. The overall structure of our image domain cGAN $\phi_{\textnormal{image}}$ in presented in Figure~\ref{fig:cgan_res_vdsr}. The number of channels used at each layer are given at the bottom of each layer output. The inputs of the generator network $\hat{x}$ are patches extracted from lower-quality reconstructions and the ground truth for training $x$ are corresponding patches extracted from full-view MBIR reconstructions \cite{jin2015model}. The goal of $\phi_{\textnormal{image}}$ is to encapsulate patch-based image priors by learning low-to-high quality image mappings. The architecture of our image domain generator network $\phi_{\textnormal{image}}$ is inspired from \cite{kim2016accurate}. It is a fully convolutional architecture and follows a residual learning strategy. It has $16$ convolutional layers with $5\times5$ kernels each performing $1$-pixel strided convolutions, resulting in a theoretical ERF of $65\times65$. The image discriminator network $D_{\textnormal{image}}$ has $5$ convolutional and $2$ fully connected layers. The convolutional layers in $D_{\textnormal{image}}$ use $3\times3$ kernels and perform $2$-pixel strided convolutions.

\subsection{Consensus Equilibrium Framework for Information Fusion}



We choose the first CE agent to correspond to a standard data fidelity term arising in MAP estimation problems:
\begin{equation}
\small
	F_{\textnormal{data}}(v_1)
	=
	\argmin_{x \geq 0} \frac{1}{2} \| \hat{y}_{\textnormal{consistent}} - \mathbf{A} x \|_W^2 + \frac{1}{2\sigma^2} \| v_1 - x \|_2^2
	\label{eq:datamap}
\end{equation}
where $W \in \mathbb{R}^{M \times M}$ is a diagonal data weighting matrix with $w_i \approx 1/var([\hat{y}_{\textnormal{consistent}}]_i)$, and $\sigma$ is the trade-off parameter between the two terms. This agent serves to incorporate our data-domain DNN information through the physical tomographic projection model $\mathbf{A}$. 

For the second CE agent we use an image domain DNN model. In particular, we choose:
\begin{equation}
	F_{\textnormal{image}}(v_2)
	=
	\phi_{\textnormal{image}}(v_2)
\end{equation}
where $\phi_{\textnormal{image}}$ is a DNN.


Following \cite{buzzard2018plug,Bouman:18} the solution of the CE equations can be obtained as follows. First define the vectorized aggregate variable maps:
\begin{equation}
	\mathbf{F}(\mathbf{v};\mathbf{\sigma}) =  
	\Bigg(
	\begin{tabular}{c}
		$F_{\textnormal{data}}(v_1;\sigma)$\\
		$F_{\textnormal{image}}(v_2)$\\
	\end{tabular}
	\Bigg)
\end{equation}


\begin{algorithm}
	\caption{DICE Algorithm for Limited Angle CT Reconstruction}
	\begin{algorithmic}[1]
		\label{algo}
		\renewcommand{\algorithmicrequire}{\textbf{Input:}}
		\renewcommand{\algorithmicensure}{\textbf{Output:}}
		\REQUIRE $y_{\textnormal{limited}}, \rho, \mathbf{\sigma}$
		\ENSURE  $x$ (reconstructed image)
		\STATE \textit{Data Domain DNN:} \\
			$\hat{y}_{\textnormal{complete}} = \phi_{\textnormal{data}}(y_{\textnormal{limited}})$ \\
		\STATE \textit{CE Initialization:} \\
		$\hat{x} = \text{FBP}(\hat{y}_{\textnormal{complete}})$ \\
		$x^{(0)} = \phi_{\textnormal{image}}(\hat{x})$ \\
		$\hat{y}_{\textnormal{consistent}} = \mathbf{A} x^{(0)}$ \\
		$\mathbf{z^{(0)}} \xleftarrow{} \left[ x^{(0)}; x^{(0)} \right]$\\
		$\mathbf{X}^{(0)} \xleftarrow{} \mathbf{G}(\mathbf{z^{(0)}}) $\\
		$k \xleftarrow{} 0$\\
		\STATE \textit{CE Solution:} \\
		\WHILE{not converged}
		\STATE $\mathbf{v} \xleftarrow{} (2 \mathbf{G} - \mathbf{I} ) \mathbf{z^{(k)}}$
		\STATE $\mathbf{X}^{(k+1)} \xleftarrow{} \mathbf{F}(\mathbf{v};\mathbf{\sigma}) $
		\STATE $\mathbf{z^{(k+1)}} \xleftarrow{} 2 \mathbf{X}^{(k+1)} - \mathbf{v} $
		\STATE $\mathbf{z^{(k+1)}} \xleftarrow{} \rho \mathbf{z^{(k+1)}} + (1-\rho) \mathbf{z^{(k)}} $
		\STATE $k \xleftarrow{} k + 1$
		\ENDWHILE
		\RETURN $x^* \xleftarrow{} \bar{\mathbf{z}}^{(k)}$ 
	\end{algorithmic} 
\end{algorithm}


Further, define the averaging or redistribution operator $G$:
\begin{equation}
	\mathbf{G}(\mathbf{v}) 
	=   
	\Bigg(
	\begin{array}{c}
	\bar{\mathbf{v}}\\
	\bar{\mathbf{v}}
	\end{array}
	\Bigg)
\end{equation}
where $\bar{\mathbf{v}} = \sum_{i=1}^2 \mu_i v_i$. 

It can be shown that $\mathbf{z}^*= x^* - \mathbf{u}^*$ is a fixed point of the map $\mathbf{T} = (2\mathbf{F}-\mathbf{I})(2\mathbf{G}-\mathbf{I})$. Once $\mathbf{z}^*$ is found the CE solution can be easily computed as $x^* = \bar{\mathbf{z}}^* = \sum_{i=1}^2 \mu_i z_i$. One way to achieve this fixed point $\mathbf{z}^*$ is using Mann iterations:
\begin{equation}
	\mathbf{z^{(k+1)}} = (1-\rho) \mathbf{z^{(k)}} +  \rho \mathbf{T} \mathbf{z^{(k)}} 
\end{equation}
for all $k \geq 0$, and $\rho \in (0,1)$, where $\mathbf{z^{(0)}}$ is an initial estimate. Finally, our DICE approach that integrates data and image domain DL using CE for CT image reconstruction is provided in Algorithm 1. In this work, we simply limit the number of outer iterations to $4$, which was observed to be adequate. Further work on convergence analysis will be conducted in the future. The application of the data agent (\ref{eq:datamap}) is accomplished by $20$ iterations of the conjugate gradient algorithm. We use $\rho=0.25, \mu_1=0.6, \mu_2=0.4$, and $\sigma^2=10^{-8}$. We use Tensorflow \cite{abadi2016tensorflow} for DL components of our approach and ASTRA toolbox \cite{van2016fast} for accelerated forward and back projection operations on GPU. \par



\section{Experiments}
We describe our experimental dataset, cGAN training strategy, and present results in this section. The real security CT dataset that we use for this work was collected using an Imatron C300 scanner as part of a data collection campaign \cite{crawford2014advances}. The scans were performed with $130 KeV$ peak source energy to image $475\textnormal{mm} \times 475\textnormal{mm}$ field of view. The acquired data was rebinned to a parallel beam geometry with $1024$ detector channels and $720$ projection angles. We split the scans into training data with $168$ bags and testing data with $21$ bags. For this work, we do not consider slices containing metallic objects for this work. The input and reference projection data was zero-padded to match the network input and output size. \par

To train the data domain cGAN we alternated between descent steps on $\phi_{\textnormal{data}}$ and $D_{\textnormal{data}}$. An additional $5-i$ gradient steps were performed on $D_{\textnormal{data}}$ for $i=1,2,3,4$ epochs  \cite{ghani2019deepmar}. We used the Adam optimizer \cite{kingma2014adam} with mini-batch size of $8$, and learning rate of $0.0002$, and momentum parameters $\beta_1=0.9, \beta_2=0.999$. We empirically selected the hyper-parameter to be $\lambda=100$. We trained the data-domain cGAN for $47$ epochs on $31,210$ examples for data completion task.\par



For the image domain cGAN, we used full-view MBIR \cite{jin2015model} reconstructions as ground truth images. We apply $40$ iterations of the conjugate gradient algorithm to perform $\ell_2$ regularized inversion using data-domain cGAN completed projection data ($\hat{y}_\textnormal{complete}$) and use them as low-quality input images. For the FBP + post-processing method, $90^0$ limited angle data was used to compute the low-quality input images (for training fairness). The image domain cGAN was trained on image patches corresponding to non-empty reference patches. We alternated between $\phi_{\textnormal{image}}$ and $D_{\textnormal{image}}$ to perform descent steps. Additional $5-i$ gradient steps were performed on $D_{\textnormal{image}}$ for $i=1,2,3,4$ epochs \cite{ghani2019deepmar}. We used the Adam optimizer \cite{kingma2014adam} with mini-batch size of $128$, and learning rate of $0.0002$, and momentum parameters $\beta_1=0.9, \beta_2=0.999$. We empirically selected the hyper-parameter to be $\lambda=10^{-5}$. We trained the image domain cGAN for $20$ epochs on $399,823$ image patches.\par




\begin{figure}
    \centering
    \includegraphics[width=0.45\textwidth]{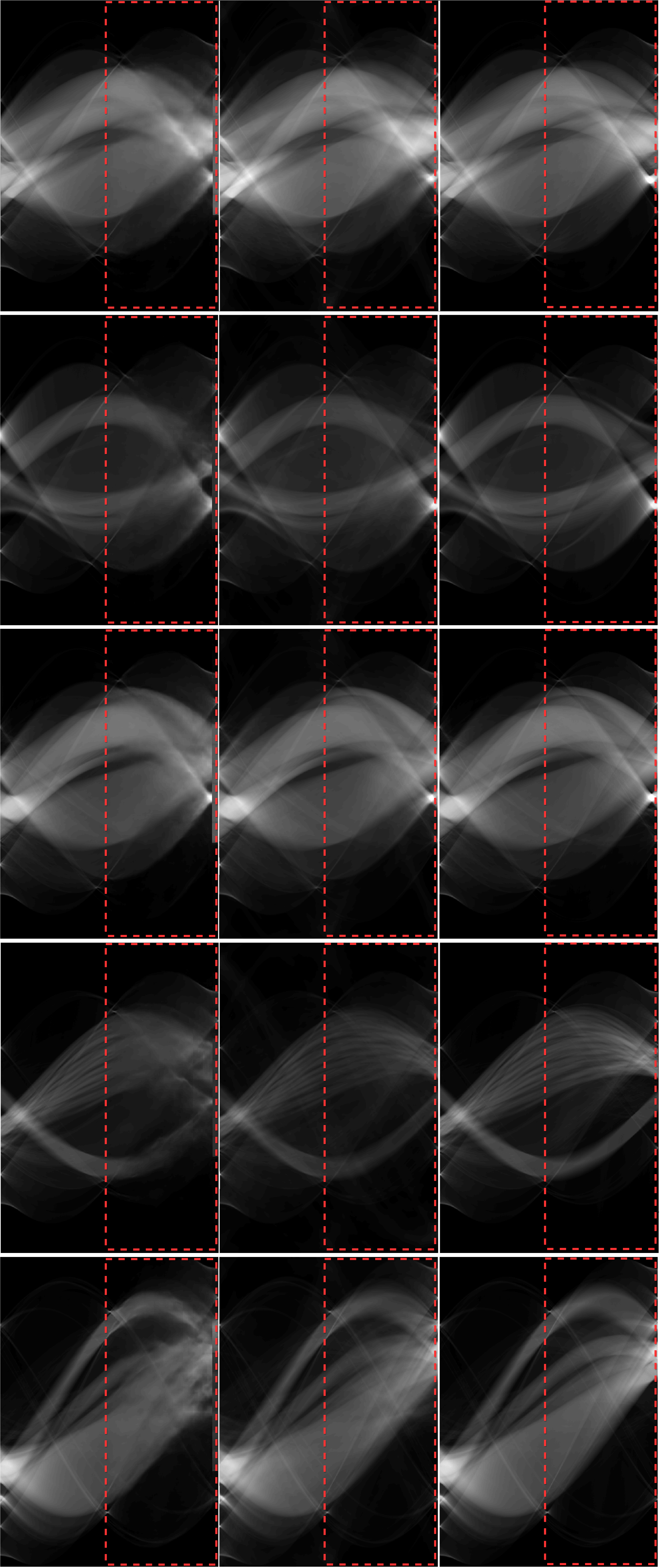}\\
    \hspace{2em}(a) $\hat{y}_{\textnormal{complete}}$  \hspace{2em}(b) $\hat{y}_{\textnormal{consistent}}$  \hspace{3em}(c) $y_{\textnormal{reference}}$ \hspace{2em}
    \caption{Data completion results are presented along with reference full-view projection data. Each row presents results for a different example. Regions corresponding to completed projection data are highlighted using red rectangles.}
    \label{fig:sino}
\end{figure}

\begin{figure*}
    \centering
    \footnotesize
	\rotatebox{90}{\hspace{2.8em} Ground Truth \hspace{6em} DICE \hspace{4.25em}  Initialization + MBIR \hspace{2.5em}  DC + FBP + PP \hspace{5em} DC + FBP \hspace{6em} FBP + PP \hspace{7.5em} FBP \hspace{2em} }
    \includegraphics[height=0.95\textheight]{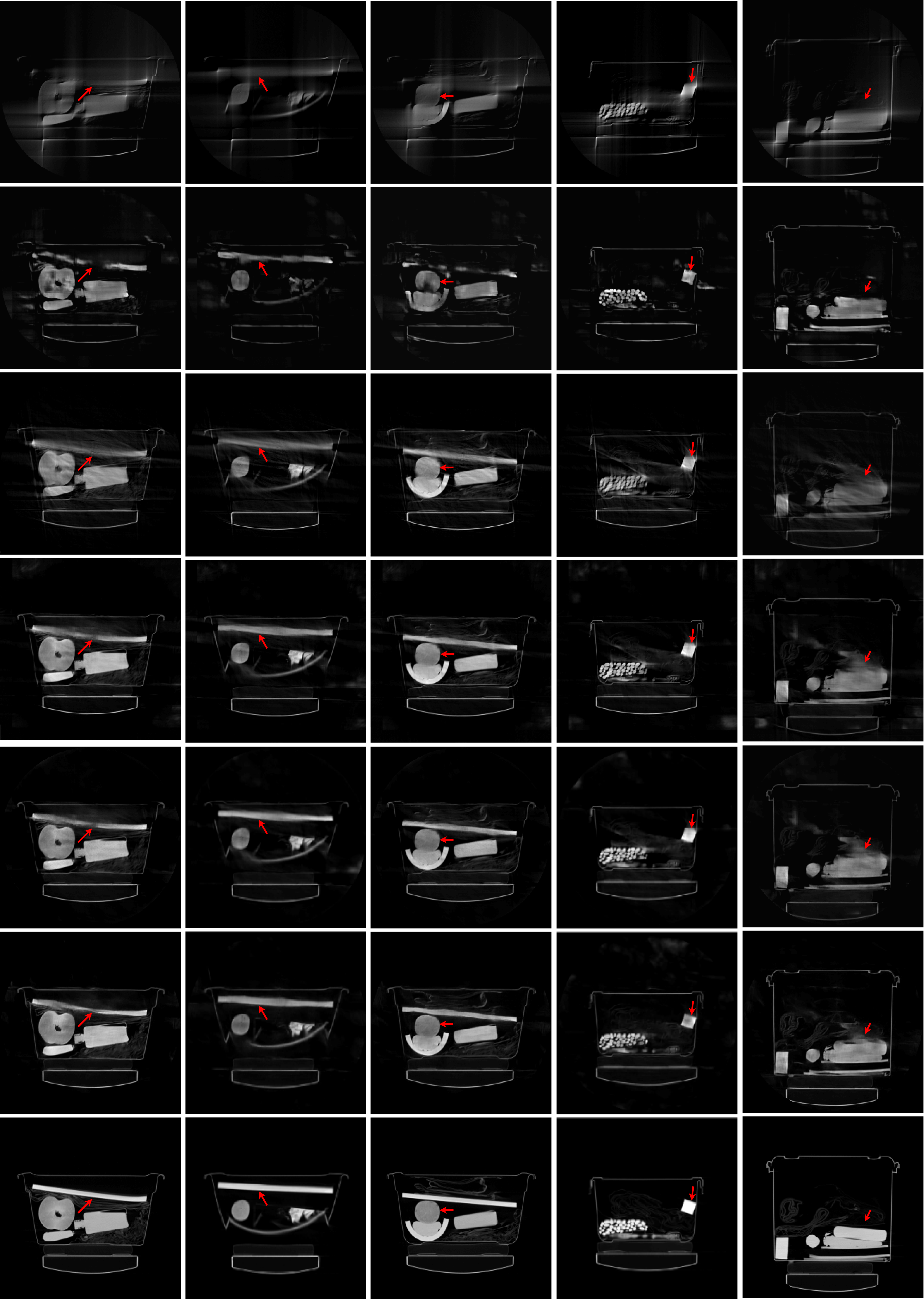}
    \caption{Limited angle CT reconstruction results using different methods are presented. Here DC refers to data domain cGAN based Data Completion, PP refers to post-processing using image domain cGAN, and Initialization refers to consistent complete data and reconstruction initialization. We use the same initialization and the same consistent completed projection data for both Initialization + MBIR and DICE methods for a fair comparison. State-of-the-art image post-processing method (second row) fails to recover lost information completely. Evidently our proposed DICE approach reconstruct superior quality images which not only recover lost information accurately but also greatly suppress image artifacts.}
    \label{fig:recons}
\end{figure*}


\subsection{Results}

First, we present data completion results along with the reference full-view projection data in Figure~\ref{fig:sino}. Column (a) presents data completion results $\hat{y}_{\textnormal{complete}}$ obtained using $\phi_{\textnormal{data}}$ network. Column (b) presents completed projection data $\hat{y}_{\textnormal{consistent}}$ that follows data consistency conditions. Reference full-view data is presented in column (c). Each row presents results for a different example. Due to imperfections in the data completion learning task and since data consistency conditions are not in place, artifacts are visible in the results presented in column (a). However, since $\hat{y}_{\textnormal{consistent}}$ follows data consistency conditions, the results in column (b) are visually more plausible. \par

Limited angle CT images reconstructed using different methods are presented in Figure~\ref{fig:recons}. Each column presents results for a different example, corresponding to the same examples considered in rows of Figure \ref{fig:sino}. The FBP reconstructions in the first row have large areas of lost structure as well as streaking artifacts. DNN post-processing of the FBP images in the second row serves to recover some of the lost structure, but fails to correct CT numbers and much structure is still mot recovered. Data domain DL aims to recover structure by learning to complete projection data prior to image reconstruction. The third row shows conventional reconstruction of this completed data. While it does a good job recovering lost structure, artifacts can still be seen. The fourth row shows post-processing of these completed data conventional reconstructions, which can be seen to result in slight improvements, such as intensity uniformity. The fifth row shows the results of combining consistent completed data and reconstruction initialization with $4$ iterations of MBIR \cite{jin2015model}, which produces minor improvements. Our proposed integrated data and image domain DNN method is shown in the sixth row. The reconstruction produces more compact object shapes consistent with the ground truth and more uniform CT numbers. The reference images are shown in the bottom row. 

Finally, we also perform quantitative analysis on $315$ slices where ground truth images are full-view MBIR reconstructions \cite{jin2015model} in Table \ref{tab:rec}. We consider three metrics RMSE: root mean square error (HU), PSNR: peak signal-to-noise ratio, and SSIM: structural similarity index. Our DICE approach involving data and image domain DL using CE framework outperforms all the considered methods on all three metrics. In our analysis, DICE reconstruction results are not heavily dependent upon initialization. Data completion alone combined with FBP significantly outperforms image post-processing approach, suggesting the value of the part of the process to the overall result.

\begin{table}[tb]
  \centering
  \caption{Reconstruction performance comparison of different methods on test dataset in terms of average RMSE (HU), PSNR (dB), and SSIM.}
    \begin{tabular}{|l|c|c|c|}
    \hline
 \textbf{Method} & \textbf{RMSE} & \textbf{PSNR} & \textbf{SSIM}\\ \hline
FBP & $137$ & $20.96$ & $0.50$ \\ \hline
FBP+PP   & $115$ & $22.46$ & $0.44$ \\ \hline
DC + FBP   & $95$ & $24.18$ & $0.73$ \\ \hline
DC + FBP +  PP & $79$ & $25.86$ & $0.72$ \\ \hline
Initialization + MBIR  & $76$ & $26.14$& $0.78$  \\ \hline
DICE (proposed)   & $\mathbf{73}$ &  $\mathbf{26.55}$ &   $\mathbf{0.81}$ \\ \hline
    \end{tabular}%
  \label{tab:rec}%
\end{table}%

\section{Conclusion}

In this work we presented DICE framework integrating both data-domain and image-domain deep learning using consensus equilibrium. Our motivation was to solve challenging limited angle CT problems. We demonstrated the approach on a limited angle security CT data set comparing the method to a variety of alternatives and showing superior results. This initial work demonstrates the potential value of combining deep learning in these two complementary domains. Ongoing work is aimed at folding our data-domain models directly into the CE framework.  

\section*{Acknowledgement}
\small
This material is based upon work supported by the U.S. Department of Homeland Security, Science and Technology Directorate, Office of University Programs, under Grant Award 2013-ST-061-ED0001. The views and conclusions contained in this document are those of the authors and should not be interpreted as necessarily representing the official policies, either expressed or implied, of the U.S. Department of Homeland Security.



\section*{Appendix 1: cGAN Objective Function}
The overall cGAN cost function consists of two terms: a conventional pixel based loss term and an adversarial loss term. Isola \etal \cite{pix2pix17} reported that using such a combination results in better performance. Generator and discriminator networks in our cGANs are jointly trained by optimizing a mini-max cost function \cite{pix2pix17,goodfellow2014generative}: 

\begin{equation}
\begin{aligned}
	\phi_i^* &= \arg \min\limits_{\phi_i} \max\limits_{D_i} \mathcal{L}_{cGAN}(\phi_i,D_i) \\
	&+ \lambda \mathbb{E}_{In,GT} [\| GT - \phi_i(In) \|_2^2]
	\end{aligned}
	\label{eq:overall}
\end{equation}

\noindent where $In$ and $GT$ represent input and ground truth used for network training, $\phi_i(In)$ represents output of $i^{th}$ generator network, $D_i$ represents $i^{th}$ discriminator network, $\lambda$ is a hyper-parameter used to control the contribution of both terms in the overall cost function,and  $\mathbb{E}_{In,GT}$ describes expectation over input and output pair dataset density. Both networks $\phi_i$ and $D_i$ play a mini-max game by acting adversaries -- $\phi_i$ efforts to output images similar to $GT$, and $D_i$ learns to differentiate between ($In,\phi_i(In)$) and ($In,GT$) pairs. This interaction is captured by the first term $\mathcal{L}_{cGAN}(\phi_i,D_i)$ of overall cost function, and is defined as: 

\begin{equation}
	\begin{aligned}
	\mathcal{L}_{cGAN}(\phi_i,D_i) &= \mathbb{E}_{In,GT} [\log D(In, GT)]\\
	& +  \mathbb{E}_{In} [\log (1 - D(In, \phi_i(In))]
	\end{aligned}
\end{equation}

\noindent where $\mathbb{E}_{In}$ represents input dataset density. We use the $\ell_2$ weight regularization on the generator network for efficient training. Additionally, we use one-sided label smoothing as suggested by Salimans \etal \cite{salimans2016improved}.

{\small
\bibliographystyle{IEEEtran}
\bibliography{literature}
}

\end{document}